\documentclass[a4paper]{jpconf}
\usepackage{graphicx}
\usepackage{cite}
\newcommand{\KV}{{\mbox{$\kappa \sigma^{2}$}}}
\newcommand{\SD}{{\mbox{$S \sigma$}}}
\newcommand{\VM}{{\mbox{$\sigma^{2}/M$}}}

\newcommand{\sNN}{{{$\sqrt{s_{_{\mathrm{NN}}}}$}}}

\newcommand{ \be }{\begin{equation}}
\newcommand{ \ee }{\end{equation}}

\begin{document}
\title{Fluctuations of Conserved Quantities in High Energy Nuclear Collisions at RHIC}

\author{Xiaofeng Luo}

\address{ Institute of Particle Physics and Key Laboratory of Quark \& Lepton Physics (MOE), Central China Normal University, Wuhan, 430079, China.}

\ead{xfluo@mail.ccnu.edu.cn}

\begin{abstract}
Fluctuations of conserved quantities in heavy-ion collisions are used to probe the phase transition and the QCD critical point for the strongly interacting hot and dense nuclear matter. The STAR experiment has carried out moment analysis of net-proton (proxy for net-baryon (B)), net-kaon (proxy for net-strangeness (S)), and net-charge (Q). These measurements are important for understanding the quantum chromodynamics phase diagram. We present the analysis techniques used in the moment analysis by the STAR experiment and discuss the moments of net-proton and net-charge distributions from the first phase of the Beam Energy Scan program at the Relativistic Heavy Ion Collider.

\end{abstract}

\section{Introduction}
One of the main goal of heavy-ion collisions is to explore the phase structure of the hot and dense nuclear matter. The phase structure can be displayed in the two dimensional Quantum Chromodynamics (QCD) phase diagram (temperature, $T$ vs. baryon chemical potential, $\mu_{B}$). It is conjectured on the basis of theoretical calculation, at finite temperature and vanishing $\mu_B$ ($\mu_B=0$) region,  there is a crossover transition between hadronic phase and partonic phase, while at large $\mu_B$ region, the phase transition is of the first order.  Thus, there should be the so called QCD Critical Point as the endpoint of the first order phase boundary towards the crossover region, which is a second order phase transition point~\cite{QCP_Prediction}. Since the {\it ab initio} Lattice QCD calculation encounters a serious so called sign problem at finite $\mu_B$ region, there are large uncertainties in determining the location of the CP or even its existence~\cite{qcp,qcp_Rajiv}.  Experimental confirmation of the existence of the CP will be an excellent verification of QCD theory in the non-perturbative region and a milestone of exploring the QCD phase structure. It is one of the main goals of the Beam Energy Scan (BES) program at the Relativistic Heavy Ion Collider (RHIC). By tuning the colliding energy of the gold nuclei from high to low values, we can vary the $T$ and $\mu_B$ of the nuclear matter created in heavy-ion collisions~\cite{bes}. This enables us to probe a broad region of the QCD phase diagram. On the other hand, fluctuations of conserved quantities, such as net-baryon (B), net-charge (Q) and net-strangeness (S),  are predicted to be sensitive to the correlation length of the system~\cite{qcp_signal,ratioCumulant,Neg_Kurtosis} and directly connected to the susceptibilities computed in the theoretical calculations~\cite{science,Lattice}. Thus, it can serve as a powerful tool to probe the phase transition and CP signal in heavy-ion collisions. Experimentally, the STAR experiment has measured the energy dependence of moments (up to  the fourth order) of event-by-event net-proton (proton minus anti-proton number, proxy of net-baryon~\cite{Hatta}) and net-charge multiplicity distributions in Au+Au collisions at {\sNN}= 7.7, 11.5, 19.6, 27, 39, 62.4 and 200 GeV.  Those data are collected from the first phase of the RHIC BES in the years 2010 and 2011.  
In the year 2014, another energy point 14.5 GeV is successfully taken and can fill in a large $\mu_B$ gap between 11.5 and 19.6 GeV. In this paper, we will present 
recent experimental results from RHIC/STAR experiment for fluctuations of conserved quantities as well as the analysis techniques. The physics implication of the results will then be discussed.

The paper is organized as follows: In the second section, we will give a brief discussion of the techniques used in the moment analysis. The experimental results and discussion will be presented in the third section. Finally, we will give a summary and outlook.

\section{Analysis Techniques}
During the last five years, a series of analysis techniques have been applied to the moments of conserved quantities distributions in heavy-ion collisions. Those include
: (1) Centrality bin width correction~\cite{WWND2011,technique}. (2) Novel centrality determination to account for the effects of centrality resolution and auto-correlation~\cite{technique}.  (3) Efficiency correction and error estimation~\cite{Delta_theory,voker_eff1,voker_eff2,Unified_Errors}. Those techniques are very important to address the background effects and extract the dynamical fluctuation signals from the observables.  In heavy-ion collisions, we cannot directly measure the collision centrality and/or initial collision geometry of the system of two nuclei. The centrality in heavy-ion collisions is generally determined by comparing the measured particle multiplicity with the Glauber Monte Carlo simulations. It is denoted as a percentage value (for e.g. 0-5\%, 5-10\%,...) for a collection of events to represent the fraction of the total cross section. This in general can cause two undesirable effects in the moment analysis of particle multiplicity distributions within finite centrality bin. One is the so called centrality bin width effect, which is caused by volume variation within a finite centrality bin size and the other one is centrality resolution effect, which is due to the initial volume fluctuations.  Let's review those techniques one by one.

\subsection{Centrality Bin Width Correction}
The centrality bin width correction is to address the volume fluctuations effects on the higher order moments of conserved quantities distributions within a finite width centrality bin. It is also called centrality bin width effect. This effect needs to be eliminated, as an artificial centrality dependence could be introduced by it. To do this, the
centrality bin width correction is applied to calculate the various moments of particle multiplicities
distributions in one wide centrality bin. Experimentally, the smallest centrality bin is determined by a
single value of particle multiplicity.
Experimental results are usually reported for a wider centrality bin (a range of particle multiplicity),
such as $0-5\%$,$5-10\%$,...etc., to reduce statistical errors. To eliminate the centrality bin width effect, we
calculate the various order cumulants ($C_{n}$) for each single particle multiplicity within one
wider centrality bin and weighted averaged by the corresponding number of events in that multiplicity.
\begin{equation} \label{eq:cbwc}
{C_n} = \frac{{\sum\limits_{r = {N_1}}^{{N_2}} {{n_r}C_n^r} }}{{\sum\limits_{r = {N_1}}^{{N_2}} {{n_r}} }} = \sum\limits_{r = {N_1}}^{{N_2}} {{\omega _r}C_n^r} 
\end{equation}
where the $n_r$ is the number of events for multiplicity value $r$ and
the corresponding weight for the multiplicity $r$, ${\omega _r} = {n_r}/\sum\limits_{r = {N_1}}^{{N_2}} {{n_r}}$. $N_1$ and $N_2$ are the lowest and highest multiplicity values for one centrality bin. Once getting the centrality bin width corrected cumulants via Eq. (\ref{eq:cbwc}), we can calculate the various order moments, for e.g. {\KV}$=C_{4}/C_{2}$ and {\SD}$=C_{3}/C_{2}$, where the $\kappa$ and $S$ are kurtosis and skewness, respectively. The final statistical error of moments for one centrality can be evaluated by standard error propagation based on Eq. (\ref{eq:cbwc}). For more details, one can see~\cite{technique}.

\subsection{Centrality Resolution and Auto-correlation Effects}
Particle multiplicity are usually used in the centrality determination, as it can reflect the initial geometry of heavy-ion collision. However, the relation
between measured particle multiplicities and collision geometry is not one-to-one correspondence and there are 
fluctuations in the particle multiplicity even for a fixed collision geometry. Thus,
one could obtain a finite resolution of initial collision geometry (centrality resolution) by using particle multiplicity to determine the centrality.
The more particles are used in the centrality determination, the better centrality resolution and smaller 
fluctuation of the initial geometry (volume fluctuation) we get. This may affect moments of the event-by-event multiplicity distributions.
On the other hand, we use the multiplicity of charged kaon and pion to define the collision centrality
in Au+Au collisions. This is to prevent the effect of auto-correlation between protons/antiprotons involved in our moments analysis and in the centrality definition.
The auto-correlation effect will results in the suppressing values of the moments. To avoid the auto-correlation effect, we should exclude the corresponding proton/antiproton from the centrality definition.  
The two background effects can be well avoided by a novel centrality definition using particle multiplicities.
For more details, one can see~\cite{technique,CPOD2013}.
\subsection{Efficiency Correction and Error Estimation}
Finite detecting efficiency for particle yield measurement in heavy-ion collisions can be easily corrected. 
However, it is not straightforward to get the efficiency corrected results for higher
moments of particle multiplicity distributions. Since the fluctuation analysis is statistics hungry, it is crucial to get the correct statistical errors with limited statistics. We provide a unified description of efficiency correction and error estimation for various order moments of multiplicity distributions~\cite{Unified_Errors}. The basic idea is to express the moments and cumulants in terms of the factorial moments, which can be easily corrected for efficiency effects. By knowing the covariance between multivariate factorial moments, we use the standard error propagation based on Delta theorem to obtain the error formula for efficiency corrected moments. This method can also be applied to the phase space efficiency case, where the efficiency of proton or anti-proton are not constant within studied phase space. One needs to note that the efficiency correction and error estimation should be done for events with the same particle multiplicity used for centrality determination and just before the centrality bin width correction.
\section{Experimental Results}
Recently, the STAR experiment has published the beam energy dependence of moments (up to forth order) of event-by-event net-proton~\cite{STAR_BES_PRL} and net-charge~\cite{netcharge_PRL} multiplicity distributions in Au+Au collisions at {\sNN}= 7.7, 11.5, 19.6, 27, 39, 62.4 and 200 GeV. For net-proton analysis, the protons and anti-protons are identified with ionization energy loss in the Time Projection Chamber (TPC) of the STAR detector within the transverse momentum range $0.4<p_{T}<0.8$ GeV/c and at mid-rapidity $|y|<0.5$.  In the net-charge case, the charged particles are measured within transverse momentum range $0.2<p_{T}<2$ GeV/c and pseudo-rapidity range $|\eta|<0.5$. 

Figure 1 shows the energy dependence of {\VM}, {\SD} and {\KV} of net-proton and net-charge distributions of Au+Au collisions for two centralities (0-5\% and 70\%-80\%) at {\sNN}= 7.7, 11.5, 19.6, 27, 39, 62.4 and 200 GeV.  The Skellam (Poisson) expectations shown in the figure reflect a system of totally uncorrelated, statistically random particle production. It predicts the {\KV} and {\SD}/Skellam to be unity for Skellam expectations as well as in the hadron resonance gas model.  For the net-proton results, the most significant deviation of {\SD} and {\KV} from Skellam distribution is observed at 19.6 and 27 GeV for 0-5\% Au+Au collisions.  At energies above 39 GeV, the results are close to Skellam expectation.  As the statistical errors are large at low energies (7.7 and 11.5 GeV), more statistics is necessary to quantitatively understand the energy dependence of {\SD} and {\KV}. To understand the effects of baryon number conservation etc., UrQMD model calculations (a transport
model which does not include a CP) for 0-5\% are presented and the results show a monotonic decrease with decreasing beam energy. For more details on baseline comparison, one can see~\cite{QM2014_baseline}. For the net-charge results,  we did not observe non-monotonic behavior for {\SD} and {\KV} within current statistics. The expectations from negative binomial distribution can better describe the net-charge data than the Poisson (Skellam) distribution. More statistics is needed for net-charge moment calculation.
\begin{figure}[htb]

\begin{minipage}[c]{0.5\linewidth}
\centering 
    \includegraphics[scale=0.33]{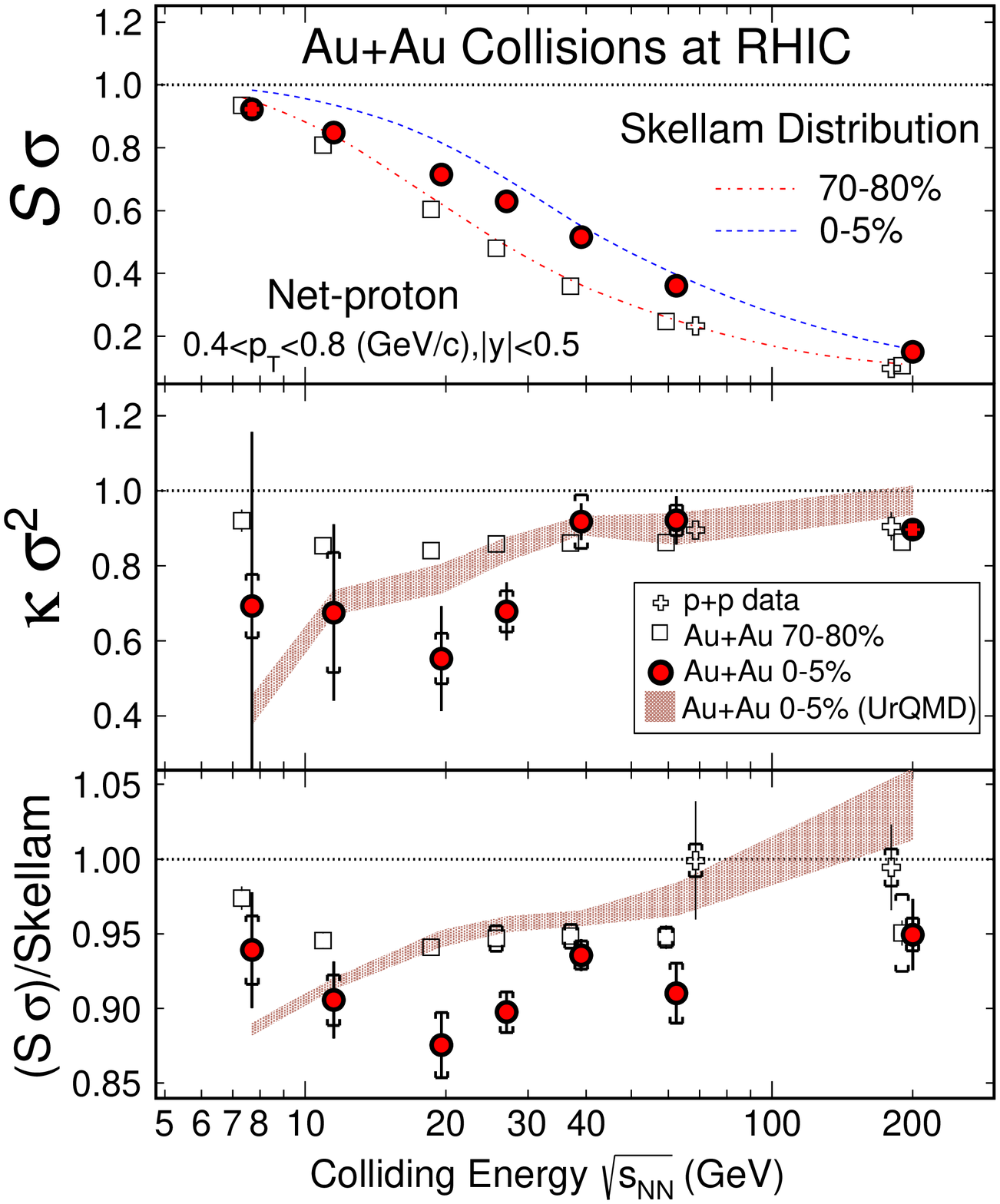}

    \end{minipage}
  \begin{minipage}[c]{0.5\linewidth}
  \centering 
   \includegraphics[scale=0.34]{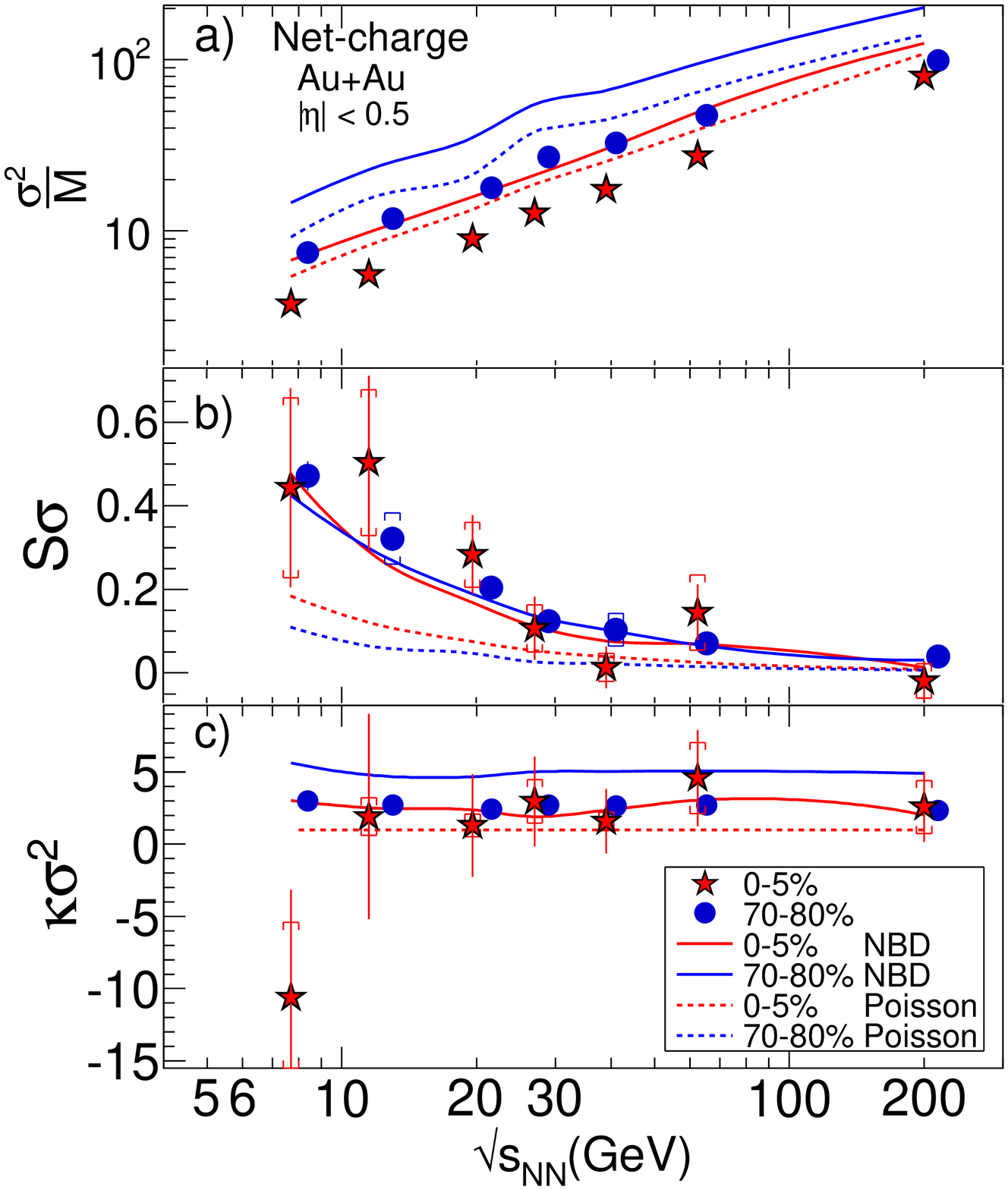}
    
      \end{minipage} 
      
     \caption{(Color online) Energy dependence of moments of net-proton (left) and net-charge (right) distributions for Au+Au collisions at RHIC BES energies.
The statistical and systematical error are shown in bars and brackets, respectively. } \label{fig:cumulants_energy}

\end{figure}

 We would like to point out: (1)  The STAR experiment has carried out moment analysis for net-proton, net-kaon, and net-charge. Different measurements are affected by kinematic cuts, resonance decays,  and other dynamical effects differently. In search for the QCD critical point,  careful studies are called for. (2) So far, the resonance decay effects are contained in the experimental results of net-proton and net-charge moments. One can estimate the decay effects by theoretical calculations and/or models. Based on the hadron resonance gas model calculation~\cite{HRG_baseline}, the decay effects for net-proton {\KV} is small and at 2\% level. But for the net-charge, the decay effects are large. (3) Based on the Delta theorem,  statistical error of cumulants ($\Delta(C_n$)) are related to the width of the distribution as $\Delta(C_n)$ $\sim$ O($\sigma^{n}$)~\cite{Delta_theory,Unified_Errors}. Thus, the wider is the distribution, the larger are statistical errors for the same number of events. That's explains why we got larger statistical errors for net-charge moments than that of net-proton, as the former has much wider distribution. (4) It is predicted that the coupling strength of pions to the critical fluctuation is smaller than protons/anti-protons. Thus, the fluctuation of net-protons will be more sensitive to the critical fluctuations than the fluctuation of net-charges (dominated by pions)~\cite{ratioCumulant, privateCom}. 

In the year 2018, the second phase of the beam energy scan program (BES-II) at RHIC will get started. During the BES-II, we will fine tune the beam energies below 20 GeV and  accumulate more events due to the increase of the luminosity of 3-10 times by using stochastic electron cooling technique. Several sub-detector upgrades for the STAR experiment are ongoing and will be ready when BES-II starts. Two of the upgrades are important for moment analysis.  The inner TPC (iTPC) upgrade will improve the tracking efficiency and enlarge the TPC acceptance. The Event Plane Detector (EPD) is 
a forward detector used to determine the event plane for Au+Au collisions. It can provide centrality determination with particles far away from the central collision region. 

\section{Summary}
In this paper, we discussed the analysis techniques used in the moment analysis and the energy dependence of moments of net-proton and net-charge distributions published by the STAR experiment. For the net-charge results, we didn't  observe non-monotonic behavior for {\SD} and {\KV}  within current statistics.  For the net-proton results, the most significant deviation of {\SD} and {\KV} from Skellam distribution is observed at 19.6 and 27 GeV for 0-5\% Au+Au collisions. More statistics are needed at low energies. In principle, one can extend the transverse momentum coverage for proton/anti-proton up to about 2$\sim$3 GeV/c with time of flight (ToF) detector at STAR for particle identification. This would allows us to have larger acceptance for protons and anti-protons in the analysis. Large acceptance is crucial for fluctuation of conserved quantities in heavy-ion collisions to probe the signals of the phase transition and the QCD critical point. 

\section*{Acknowledgement}
The work was supported in part by the MoST of China 973-Project No. 2015CB856901, NSFC under grant No. 11205067, 11221504 and 11228513.

\section*{References}
\bibliography{FAIRNESS2014}
\bibliographystyle{unsrt}

\end{document}